\documentclass[prl,showpacs,showkeys,twocolumn]{revtex4}
\usepackage{graphicx}
\begin{document}

\title{
   Novel Families of Fractional Quantum Hall States: \\
   Pairing of Composite Fermions}

\author{
   John J. Quinn$^1$, 
   Arkadiusz Wojs$^{1,2}$, 
   and Kyung-Soo Yi$^{1,3}$
   }

\affiliation{
   $^1$University of Tennessee, Knoxville, Tennessee 37996, USA\\
   $^2$Wroclaw University of Technology, 50-370 Wroclaw, Poland\\
   $^3$Pusan National University, Pusan 609-735, Korea}

\begin{abstract}
Fractional quantum Hall (FQH) states have recently been observed at 
unexpected values of the filling factor $\nu$.
Here we interpret these states as a novel family of FQH states involving 
pairing correlations rather than Laughlin correlations among the 
quasiparticles (QP's).
The correlations depend upon the behavior of the QP--QP pseudopotential 
$V_{\rm QP}(L')$, the interaction energy of a pair as a function of the 
pair angular momentum $L'$.
This behavior, known from numerical studies of small systems, is used 
to demonstrate that pairing correlations give rise to FQH states at the 
experimentally observed values of $\nu$.
\end{abstract}



\maketitle

Recently Pan {\sl et al.} \cite{pan} observed fractional quantum Hall minima 
in $\rho_{xx}$ at unexpected values of the Landau level filling factor $\nu$ 
outside the Jain sequence of states with $\nu=n(2pn \pm 1)^{-1}$, where $n$ 
and $p$ are positive integers.
The composite Fermion hierarchy \cite{sitko1,haldane}, in which the reapplication 
of the Chern--Simons (CS) mean field approximation attaches additional flux quanta 
to the quasiparticles (QP's) in a partially filled composite Fermion (CF) shell, 
has been suggested \cite{pan,smet} as a possible explanation of some of these states.
However, the form of the residual interactions \cite{sitko2,wojs1,quinn1,lee} 
between QP's sometimes precludes Laughlin correlations and the realization of 
certain daughter states of the CF hierarchy \cite{quinn2,wojs2}.
Among these states are the $\nu=4/11$ and 4/13 daughter states observed 
experimentally \cite{pan}, which correspond to quasielectron (QE) and quasihole 
(QH) filling factors $\nu_{\rm QE}=1/3$ and $\nu_{\rm QH}=1/5$, respectively.
In addition, the observed even denominator fractional fillings cannot arise 
within the CF hierarchy.
Here, we show that instead of having Laughlin correlations, the QP's form pairs, 
and that these pair excitations cause the novel incompressible daughter states 
at the unexpected values of $\nu$ observed experimentally.

By Laughlin correlations \cite{laughlin} among interacting Fermions confined to 
a spherical surface we mean that pair states with the largest values of the pair 
angular momentum $L'$ (or smallest values of the relative angular momentum 
$\mathcal{R}=2l-L'$, where $l$ is the single Fermion angular momentum) are maximally 
avoided.
Laughlin correlations occur if and {\sl only if} the pseudopotential $V(L')$ is 
``superharmonic,'' that is, rises with increasing $L'$ faster than $L'(L'+1)$ as 
the avoided value of $L'$ is approached \cite{wojs1,quinn1,quinn2,wojs3}.
Here $V(L')$ is the interaction energy of a pair of Fermions as a function of $L'$.
The pseudopotential for electrons in the lowest Landau level ($n=0$) is superharmonic 
at all values of $L'$.  
However, for the first excited Landau level ($n=1$) it is not superharmonic at the 
largest value of $L'$, namely $L'=2l-1$.
This is known to result in pairing of the electrons \cite{wojs3,wojs4} at filling 
factors $\nu=7/3$, $5/2$, and $8/3$.

\begin{figure}
\resizebox{3.2in}{1.6in}{\includegraphics{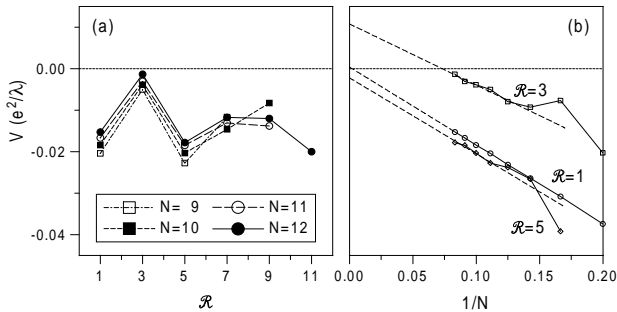}}
\caption{
   (a) 
   Interaction pseudopotentials $V_{\rm QE}(\mathcal{R})$ for a pair 
   of QE's of the Laughlin $\nu=1/3$ state, calculated for up to $N=12$ 
   electrons on a spherical surface.
   (b) 
   The leading QE pseudopotential parameters $V_{\rm QE}(\mathcal{R})$ 
   for $\mathcal{R}=1$, 3, and 5, plotted as a function of $N^{-1}$, inverse 
   of the particle number.
   Extrapolation to $N^{-1}\rightarrow 0$ corresponds to an infinite planar system.
   \label{fig1}} 
\end{figure}

For QP's of the Laughlin $\nu=1/3$ state, $V_{\rm QP}(L')$ has been obtained from 
exact numerical diagonalization studies of small systems 
\cite{sitko2,wojs1,quinn1,lee,wojs2} (with particle number $N\leq12$).
This is illustrated for QE's in Fig.~\ref{fig1}a.  
When plotted as a function of $N^{-1}$, $V_{\rm QE}(\mathcal{R})$ converges
to a rather well defined limit as shown in Fig.~\ref{fig1}b for $\mathcal{R}=1$, 
3, and 5.  
The results are quite accurate up to an overall constant (which does not affect the 
correlations).
Because the short-range interactions (i.e., at small values of $\mathcal{R}$ or small 
QE separation) determine the nature of the ground state, numerical results for small 
systems describe the essential correlations quite well for systems of any size.
Furthermore, the subharmonic character of $V_{\rm QP}$ at certain values of 
$\mathcal{R}$ ($\mathcal{R}=1$ for QE's and $\mathcal{R}=3$ for QH's) makes it 
impossible for $\nu_{\rm QE}=1/3$ and $\nu_{\rm QH}=1/5$ to lead to incompressible 
daughter states of the CF hierarchy.
By this we mean that for a spin polarized state in which QP's of the Laughlin 
$\nu=1/3$ state yield filling factors $\nu_{\rm QE}=1/3$ (or $\nu_{\rm QH}=1/5$), 
Laughlin correlations among the QP's, giving rise to incompressible daughter states 
at $\nu=4/11$ (or $\nu=4/13$), {\sl cannot} occur \cite{wojs1,quinn1,quinn2,wojs2}.

How then can we understand the novel states observed at $\nu=5/13$, 3/8, 4/11, etc.?
To illustrate, we use the case of QE's (the application to QH's is straightforward 
and will be given elsewhere).
It is apparent that $V_{\rm QE} (\mathcal{R})$ is not superharmonic at 
$\mathcal{R}=1$.
In fact, its maximum repulsion occurs at $\mathcal{R}=3$.
In this case, the QE's do not display Laughlin correlations by avoiding the pair 
state with $\mathcal{R}=1$ (or, in the planar geometry, by having the 
Laughlin--Jastrow factor $\prod_{i<j}(z_i-z_j)^2$ in the ground state wavefunction 
where $z_j=x_j-iy_j$ is the position of the $j^{th}$ QE).
Instead, they tend to form pairs with $\mathcal{R}=1$ in order to minimize the 
pair amplitude \cite{wojs3,wojs4} with $\mathcal{R}=3$.
Pairing at filling factor $\nu=5/2$ has been considered by others 
\cite{moore,greiter} based on the observation of an unexpected incompressible 
ground state.

The pairs can be thought of either as Bosons or Fermions with pair angular momentum 
$l_{\rm P}=2l-1$, since in two-dimensional systems Bosons (Fermions) can be 
transformed into Fermions (Bosons) by a CS transformation \cite{quinn3,wilczek}.
For a system containing more than a single pair, {\sl the allowed values of the 
total angular momentum of two pairs must be chosen in such a way that the Pauli 
principle is not violated when accounting for identical constituent Fermions 
belonging to different pairs}. 
This can be accomplished if different pairs are not allowed to approach too closely 
by requiring that the largest allowed value of the total angular momentum of two 
pairs (treated as Fermions) to be given by $\tilde L'=2 l_{\rm FP}$, where 
\begin{equation}
   2l_{\rm FP}=2(2l-1)-\gamma_{\rm F}(N_{\rm P}-1).
\end{equation}
Here the simplest assumption of complete pairing is made, so that $N_{\rm P}=N/2$ 
is the number of Fermion pairs (FP's), and $\gamma_{\rm F}$ will be an odd integer 
(if the pairs were treated as Bosons $\gamma_{\rm B}$ would equal $\gamma_{\rm F}-1$).
$\gamma_{\rm F}$ is chosen so that the Fermion pair filling factor 
$\nu_{\rm FP}=(N_{\rm P}-1)/2l_{\rm FP}$ is equal to unity when the QE filling 
factor $\nu_{\rm QE}=(N-1)/2l$ is also equal to unity.
This condition gives $l_{\rm FP}=2l-1-{3\over2}(N_{\rm P}-1)$ for the ``effective'' 
angular momentum of one FP and
\begin{equation}
   \nu_{\rm FP}^{-1}=4\nu_{\rm QE}^{-1}-3
\label{nuFP}
\end{equation}
for large systems.
This CS transformation automatically forbids states of two FP's with the smallest 
separation.
The larger pair--pair separation causes the constituent QE's to avoid the largest 
repulsion at $\mathcal{R}=3$.
In addition, the transformation selects from $\mathcal{D}_L(N,l)$, the number of 
multiplets of total angular momentum $L$ formed from $N$ Fermions each with angular 
momentum $l$, a subset $\mathcal{D}_L(N_{\rm P}, l_{\rm FP})$.

We expect pair formation for QE filling factor satisfying $2/3\geq\nu_{\rm QE}\geq1/3$, 
where Laughlin--Jain states \cite{laughlin,jain} that avoid $\mathcal{R}=1$ would 
normally occur for a superharmonic potential.
If we assume the FP's support Laughlin correlations, when their separations are large, 
then incompressible ground states would be expected at the Laughlin filling factors 
$\nu_{\rm FP}=1/5$, 1/7, and 1/9.
From Eq.(\ref{nuFP}), these correspond to values of $\nu_{\rm QE}$ given by 1/2, 2/5, 
and 1/3.
For $\nu_{\rm QE} \geq 1/2$ we make use of QE--QH symmetry and think of Fermion pairs 
of QH's giving $\nu_{\rm QH}=1/2$, 2/5, 1/3 and $\nu_{\rm QE}=1/2$, 3/5, and 2/3.
Only at these values of $\nu_{\rm FP}$ do we have Laughlin states of the Fermion pairs 
with $\nu_{\rm QE}$ in the required range.
In the hierarchy scheme \cite{sitko1} describing partially filled CF levels, 
the original electron filling factor is given by
\begin{equation}
   \nu^{-1}=2+(1+\nu_{\rm QE})^{-1}.
\end{equation}
Here the 2 on the right hand side comes from the addition of two flux quanta per 
electron in the original CS transformation on the electrons, and $(1+\nu_{\rm QE})$ 
is the CF filling factor.
Hence, we find new incompressible states at $\nu=5/13$, 8/21, 3/8, 7/19, and 4/11.
All except the 7/19 and 8/21 states have been observed.
We are uncertain whether there is some reason why these two states do not occur, 
or if, because they have such large denominators, they are simply difficult to see 
and might be observed in future experiments.
It could be that our simple model, which assumes complete pairing of all QP's, 
is not valid at every value of $\nu_{\rm QP}$ (alternatively, partially paired
states could be considered), or that a spin unpolarized state preempts pair 
formation.
\begin{figure}
\resizebox{3.2in}{1.6in}{\includegraphics{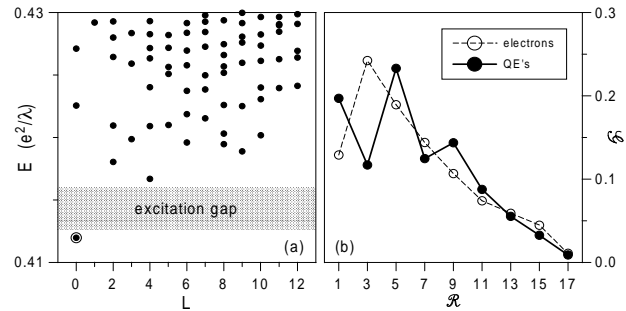}}
\caption{
   (a) 
   Low energy spectrum (energy $E$ as a function of total angular momentum $L$) 
   of 10 QE's at $2l=17$ corresponding to $\nu_{\rm QE}=1/2$ and $\nu=3/8$, 
   obtained in exact diagonalization in terms of individual QE's interacting 
   through the pseudopotential shown in Fig.~\protect\ref{fig1}.   
   $\lambda$ is the magnetic length.
   The open circle at $L=0$ denotes a robust ground state.
   (b) 
   Coefficient $\mathcal{G}(\mathcal{R})$, the amplitude associated with pair 
   states of relative pair angular momentum $\mathcal{R}$, for the lowest $L=0$ 
   state of 10 particles in a shell of angular momentum $l=17/2$.
   The solid dots and open circles are obtained for the QE's (the $\nu=1/2$ 
   ground state marked in frame a) and electrons, respectively.
   \label{fig2}}
\end{figure}

As an illustration we have performed an exact diagonalization on a system 
containing $N=10$ QE's at $l=17/2$.
This corresponds to $\nu_{\rm QE}=1/2$.  
The energy spectrum (Fig.~\ref{fig2}a) obtained using the QE--QE pseudopotential 
given in Fig.~\ref{fig1} shows a robust $L=0$ ground state and an excitation gap.
The coefficient $\mathcal{G}(\mathcal{R})=\sum_{L' \alpha'} |G_{L' \alpha'}
(\mathcal{R})|^2$ appropriate for the $L=0$ ground state is shown in Fig.~\ref{fig2}b 
for the allowed values of $\mathcal{R}$.
$G_{L' \alpha'} (\mathcal{R})$ is the coefficient of fractional grandparentage 
\cite{sitko2,wojs1,quinn1,shalit}, and $\mathcal{G}(\mathcal{R})$ is a measure 
of the amplitude associated with pair states of relative pair angular momentum 
$\mathcal{R}$.
For the purpose of comparison, the same results are shown for the $L=0$ state of 
a ten-electron system in the lowest Landau level, also at $l=17/2$.
The large decrease in $\mathcal{G}(\mathcal{R})$ at $\mathcal{R}=3$ and increase 
at $\mathcal{R}=1$ in going from $V_e(\mathcal{R})$ to $V_{\rm QE}(\mathcal{R})$ 
is clear evidence of the avoidance of pair states with $\mathcal{R}=3$ and the 
formation of pairs with $\mathcal{R}=1$ in the case of QE's.

Rather than contradicting the assertion \cite{wojs2} that the 4/11 state (or 
$\nu=4/13$ state when $\nu_{\rm QH}=1/5$) cannot occur as an incompressible daughter 
state in the standard CF hierarchy of Laughlin correlated QP states of a spin 
polarized system, the results of Pan {\sl et al.} \cite{pan} offer support for 
the idea of pairing of QP's at certain values of $\nu_{\rm QP}$.
In contrast to the hierarchies of Laughlin correlated QP states, the pairing 
picture can account for incompressible states in the lowest Landau level with 
even denominator fractional filling.
We emphasize that simple repetition of Laughlin correlations among daughter states 
\cite{sitko1,smet,mani} containing CF QP's is not always appropriate \cite{wojs2} 
because of the form of $V_{\rm QP}(\mathcal{R})$.
The proposed pairing of CF QP's gives rise to a new type of QP which in turn leads 
to a myriad of completely novel hierarchy states.

\noindent
The authors gratefully acknowledge the support by Grant DE-FG 02-97ER45657 of 
the Material Research Program of Basic Energy Sciences--US Department of Energy.
AW acknowledges support from Grant 2P03B02424 of the Polish KBN, 
and KSY acknowledges partial support of the ABRL(R14-2002-029-01002-0) through the KOSEF.

\end{document}